\documentclass[aps,prl,twocolumn,showpacs,eqsecnum,nofootinbib,floatfix]{revtex4-1}

\pdfoutput=1
\usepackage{amssymb}
\usepackage{gensymb}
\usepackage{graphicx}
\usepackage{amsmath}
\usepackage{wasysym}
\usepackage{verbatim}
\usepackage{dcolumn}   
\usepackage{bm}        

\def\aap{Astron.\ \& Astrophys.}

\usepackage{chngcntr}
\counterwithout{equation}{section} 

\begin{document}
\title{The TeV Cosmic-Ray Anisotropy from Local Dark Matter Annihilation}
\author{J. Patrick Harding}
\email{jpharding@lanl.gov}
\affiliation{Physics Division, Los Alamos National Laboratory, Los Alamos, NM, USA}
\date{\today}

\begin{abstract}
Several experiments have reported regions in the TeV sky with an excess of cosmic rays. Here we consider the possibility that the excess cosmic rays are coming from dark matter annihilation in nearby subhalos. We provide motivation for dark matter as the source of the excess and show that dark matter annihilations can naturally produce the cosmic ray anisotropy without the need for exotic dark matter models. We show that viable dark matter explanations of the cosmic ray excess are consistent with current measurements of antiprotons, positrons, and gamma-rays. Additionally, we show that the dark matter interpretation of the observed anisotropy predicts detections in several cosmic ray channels by the next generation of experiments.

\end{abstract}




\maketitle
{\parindent0pt\it Introduction. }
Large- and small-scale anisotropies in the arrival direction of TeV cosmic rays (CRs) have been observed by multiple experiments~\cite{Amenomori:2005dy,Abdo:2008kr,Aartsen:2012ma,{DiSciascio:2013cia}}. However, the source of the anisotropic CRs remains a mystery. The anisotropy is mostly hadronic, yet the gyroradius of a TeV proton is $\sim0.005{\rm\ pc}$ and the decay length of a neutron is only $\sim0.1{\rm\ pc}$, both of which are much closer than the nearest star~\cite{Abdo:2008kr}. Additionally, the spectrum of the brightest region of excess CRs, Region A, has a hard spectrum with a cutoff above $\sim10{\rm\ TeV}$, which is very different from the soft power-law spectrum of the isotropic CRs~\cite{Abdo:2008kr}. This observed anisotropy is at relatively small $\sim10\degree$ scales, so its source is expected to be a localized beam of hadrons, not merely a dipole anisotropy from standard CR diffusion. 

Production of a localized hadronic region of CRs requires non-standard CR propagation. Several sources of the CR excess have been considered, including strong-scattering of local protons~\cite{Salvati:2008dx,Drury:2008ns}, neutron emission from nearby gas in the heliosphere~\cite{Salvati:2008dx,Drury:2008ns,Lazarian:2010sq,Desiati:2012pf}, and multiple supernova remnants with stochastic particle propagation~\cite{Erlykin:2006ri,Blasi:2011fm,Pohl:2012xs}.
However, such models either require an implausibly local unknown CR source, underproduce the flux of the observed anisotropy,  or produce only a large-scale anisotropy rather than the observed small-scale anisotropic regions.

A nearby magnetic stream propagating the hadrons from a source tens to hundreds of parsecs away explains the small angular scales of the measured emission~\cite{Salvati:2008dx,Drury:2008ns}. Also, the possibility of there being such a source is greatly increased. Small structures in the Galactic magnetic field are expected~\cite{Giacinti:2011mz,Pakmor:2012xy,Xu:2013ppa}, and CR propagation through such structures tends to collimate the source CRs rather than diffusing them~\cite{Beresnyak:2010yq,2011A&A...527A..79B,Desiati:2011xg,Kistler:2012ag,Beresnyak:2013ria}, producing the observed small-scale anisotropy. Additionally, a possible bending of the coherent magnetic field from the source to the Sun naturally explains the lack of a coincident gamma-ray signal at the location of the CR excess. Though most models consider supernova remnants as the source of CRs for the stream, here we show that Dark matter (DM) annihilation in a local subhalo is a strong candidate to provide the primary hadrons for a magnetic stream.

In this letter, we discuss the consistency of the TeV CR anisotropy with a nearby DM subhalo. We propose a physically-motivated DM spectrum with a realistic subhalo spatial distribution. We discuss multiple constraints on the DM CR source and show that the proposed DM model is consistent with all known observations. Finally, we make the case that the DM model for the CR anisotropy is a strongly motivated source candidate and outline how future experiments can verify this model.

{\parindent0pt\it The Dark Matter Cosmic-Ray Spectrum. }
A DM model for the CR excess is motivated by two factors. First, in order to produce an anisotropy of such small angular scale, a source needs to be near a magnetic stream within $100 {\rm\ pc}$ of the Sun, preferably much closer. Most sources of TeV gamma-rays are known to lie much too far from the Sun to produce this behavior. Even the nearby Geminga pulsar, which has been considered as an explanation of the excess~\cite{Salvati:2008dx,2010A&A...513A..28S}, is over 250 pc away. However, small DM subhalos can naturally be much closer than $100 {\rm\ pc}$ to the Sun. Second, the spectrum of the anisotropic hadrons has a hard ($\gamma>-2$) spectrum with a cutoff energy of $\sim10{\rm\ TeV}$, which is the CR spectrum expected from DM annihilation - all canonical DM annihilation channels produce CR spectra much harder than standard astrophysical sources, with cutoff energy about an order of magnitude below the DM mass. The leptonic channels are in fact so hard that they fail to reproduce the measured spectrum of the CR anisotropy, so we do not consider leptophilic DM in this letter. 

\begin{figure}[t]
\begin{center}$
\begin{array}{c}
\includegraphics[width=3.2in]{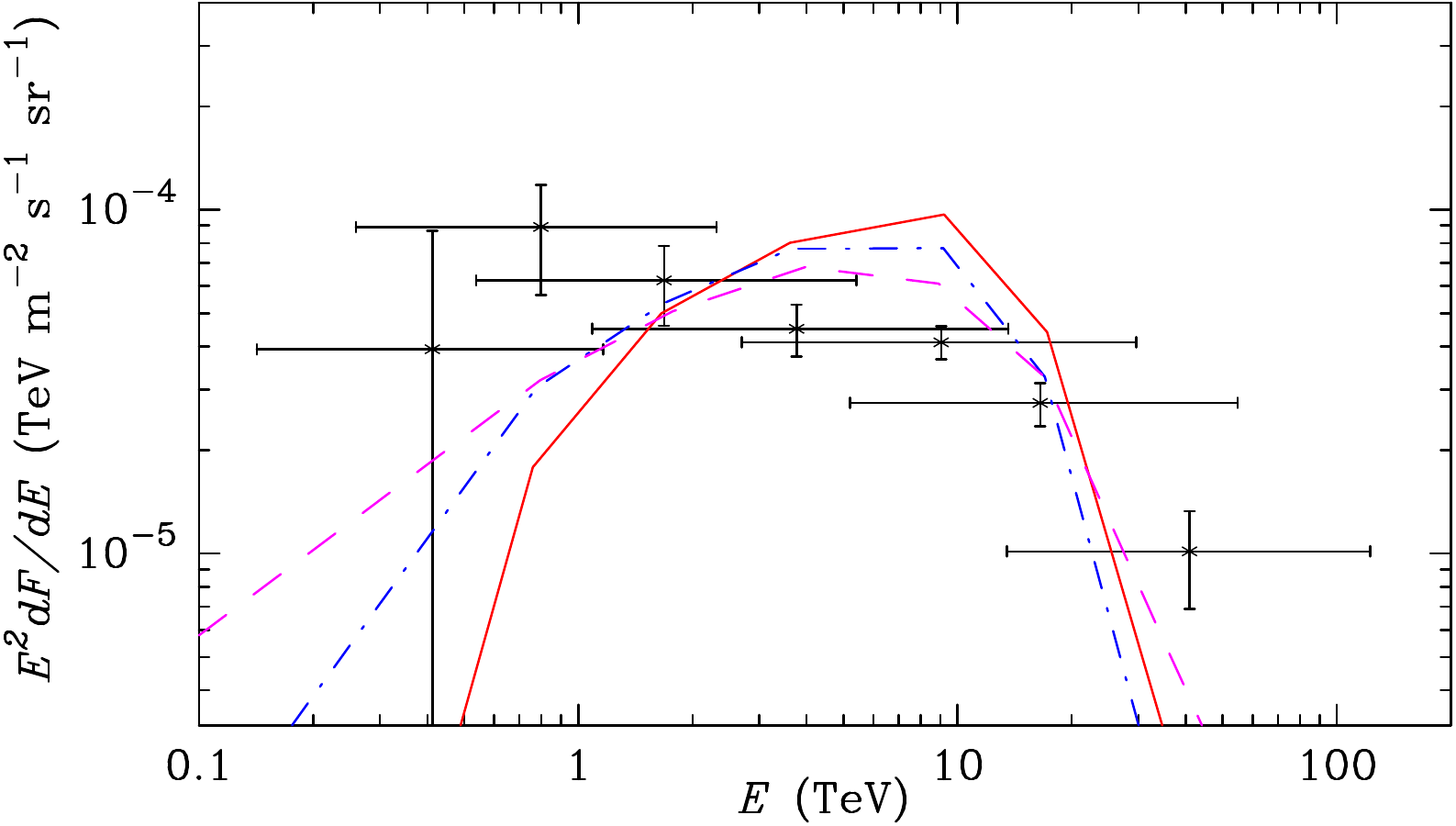}
\end{array}$
\end{center}
\caption[Three DM Antiproton Spectra Compared to data.]{\small The spectral dependence of the CR anisotropy from Region A~\cite{Abdo:2008kr}, is shown as the starred data points. To convert from relative intensity to CR flux, we have used the CR flux given in Ref.~\cite{Beringer:1900zz}. We show three candidate DM antiproton spectra, normalized to the Milagro data: a $M_\chi=60 {\rm\ TeV}$ $W^+W^-$ spectrum (solid red line), a $M_\chi=50 {\rm\ TeV}$ $Z^0Z^0$ spectrum (dot-dashed blue line), and a $M_\chi=100 {\rm\ TeV}$ $b\bar{b}$ spectrum (dashed magenta line).
\label{DMCRspectra}}
\end{figure}

A wide range of DM masses and spectra fit the CR anisotropy.
The spectral data of Region A of Ref.~\cite{Abdo:2008kr} is consistent with $W^+W^-$ and $Z^0Z^0$ annihilation channels with DM masses from $\sim30 {\rm\ TeV}$ to $\sim100 {\rm\ TeV}$ and the $b\bar{b}$ channel from $\sim40 {\rm\ TeV}$ to $\sim200 {\rm\ TeV}$.
To calculate the CR spectrum for a particular DM annihilation channel, we use {\sc PYTHIA 6.4} to simulate the decay of the annihilation products~\cite{Sjostrand:2006za}, following the method of Ref.~\cite{Abazajian:2011ak}. For hadronic CRs, we consider both proton and antiproton fluxes, and we assume that all neutrons are fully decayed near to the source.
Three of these spectra are shown in figure~\ref{DMCRspectra}: a $M_\chi=60 {\rm\ TeV}$ $W^+W^-$ spectrum, a $M_\chi=50 {\rm\ TeV}$ $Z^0Z^0$ spectrum, and a $M_\chi=100 {\rm\ TeV}$ $b\bar{b}$ spectrum. All three spectra peak between 4-10 TeV, consistent with the measured energy cutoff. Furthermore, energy losses during the CR propagation would shift the peaks to slightly lower energy and soften the cutoffs, as seen in the data. The primary difference between the spectral shapes is at low energy, but a measurement with better energy resolution could easily differentiate between these DM spectra. 

{\parindent0pt\it The Dark Matter Spatial Distribution. }
For a DM region with given density profile, the differential CR flux per solid angle is dependent on both the distance to the source and the integral over the density-squared (the ``J-factor''), as well as the DM mass and spectrum (see Ref.~\cite{Abazajian:2011ak} for details). In the Milagro analysis, a conservative $10\degree\times10\degree$ ($0.03 {\rm\ sr}$) region was considered for Region A, so the large extension of DM subhalos makes them a likely source for the CRs. It should also be mentioned that the exact configuration of the magnetic field region can have either a focusing or defocusing effect on the observed spatial extent of the source, though we neglect such fine-tuning of the exact shape of the magnetic field region to keep our conclusions robust.

Just as we considered a range of DM masses, we also consider a range of DM source subhalos. Here we consider the Navarro-Frenk-White DM density profile~\cite{Navarro:1996gj}. We calculate the profile parameters for DM subhalos from Ref.~\cite{Klypin:2010qw}, getting the J-factors shown in Table~\ref{DMJtable}. The subhalos have been placed at the closest distance $D_{\rm min}$ to the Sun which is consistent with the local DM density of $0.3\pm0.1 {\rm\ GeV\,cm^{-3}}$~\cite{Bovy:2012tw}. Over many orders of magnitude in subhalo mass, the J-factor changes minimally, so our calculations are largely independent on the choice of DM subhalo mass. However, for charged CRs the calculated J-factor is dependent not on the distance to the Sun, but on the distance to the magnetic stream, which can be very small for subhalos located less than $100 {\rm\ pc}$ away. We show this effective J-factor to the magnetic stream in Table~\ref{DMJtable} as well.
\begin{table}
\begin{center}
\begin{tabular}[t]{|c|c|c|c|}
  \hline
  $M_{\rm vir}\ (M_{\astrosun})$ & $D_{\rm min}\ ({\rm pc})$ & $J_{\Delta\Omega}(D_{\rm min})$ & $J_{\Delta\Omega}(D_{\rm min}-100{\rm\ pc})$\\
  \hline
  $10^{9}$ & 933 & 119 & 137\\
  $10^{8}$ & 465 & 114 & 158\\
  $10^{7}$ & 225 & 112 & 247\\
  $10^{6}$ & 108 & 112 & 2840\\
  $10^{5}$ & 51.3 & 111 & -\\
  $10^{4}$ & 24.1 & 110 & -\\
  $10^{3}$ & 11.2 & 109 & -\\
  \hline
\end{tabular}
\caption[Table of the J-factors for a range of DM subhalos.]{\label{DMJtable}: Table of the J-factors for a range of virial masses of DM subhalos. The distances $D_{\rm min}$ are taken such that the DM density at the Sun is $0.4 {\rm\ GeV\,cm^{-3}}$. The J-factors (given in ${\rm GeV^2\,cm^{-6}\,kpc}$) are calculated over the solid angle of $0.03 {\rm\ sr}$ for an object at a distance $D_{\rm min}$ to the Sun and for an object a distance $D_{\rm min} - 100{\rm\ pc}$ from a magnetic stream.}
\end{center}
\end{table}

In this letter, we consider a nearby subhalo with virial mass of $10^6M_{\astrosun}$. This is about an order of magnitude smaller in mass than the smallest known dwarf spheroidal galaxies but is large enough that it is not strongly sensitive to the extrapolation of simulations to low-mass subhalos. Additionally, the minimal distance to the Sun for such a subhalo is only $108 {\rm\ pc}$, which is close to the maximal length of a coherent local magnetic field while conservatively not being directly inside the magnetic region itself. However, we could consider a smaller subhalo with little loss of generality; the J-factor for a smaller subhalo should be similar. However, a smaller subhalo could be closer to the Sun, increasing the likelihood of a magnetic stream connecting the Sun to the source. 
\begin{figure*}
\begin{center}
$\begin{array}{ccc}
\includegraphics[width=0.3\textwidth]{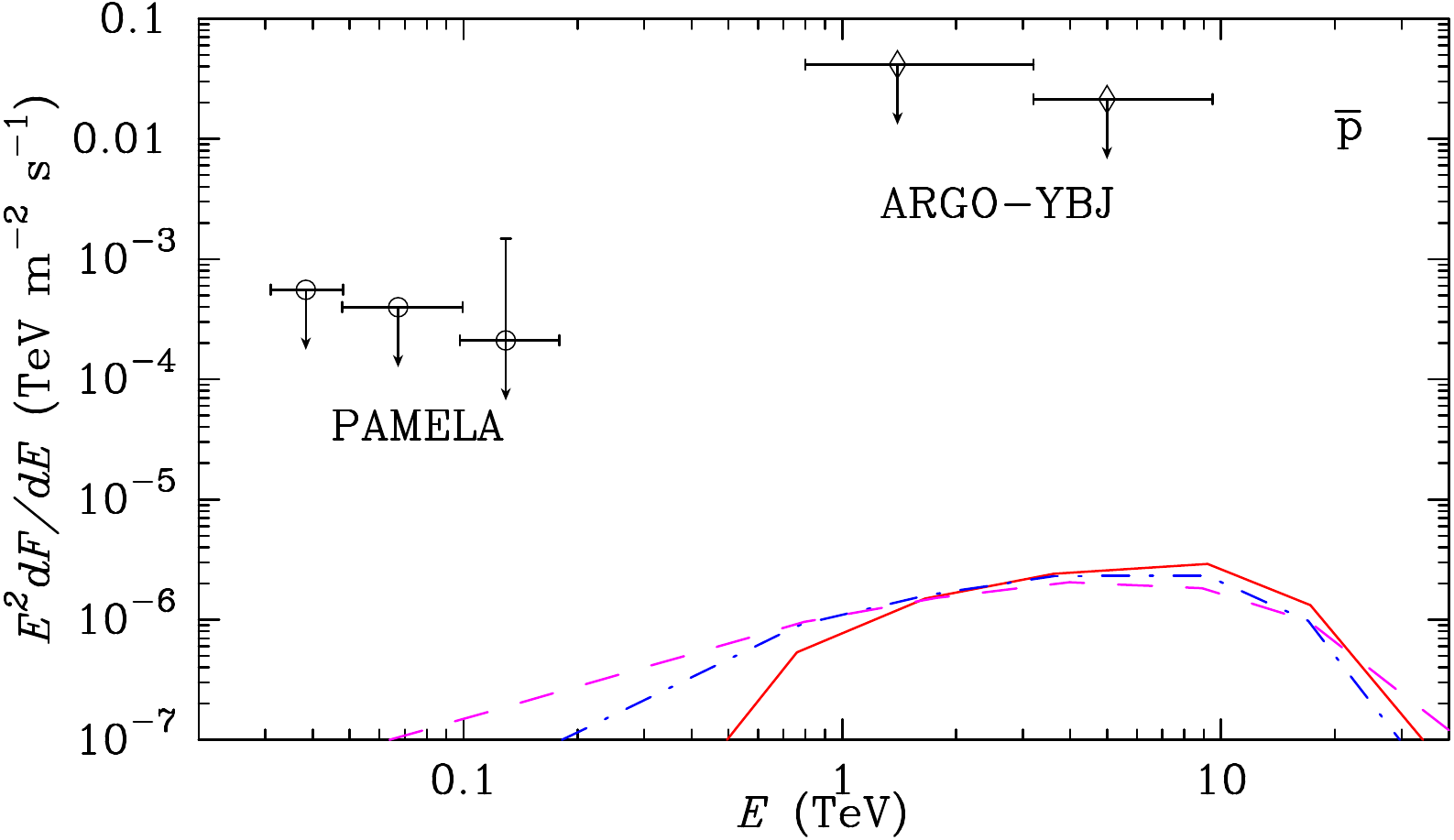} & \includegraphics[width=0.3\textwidth]{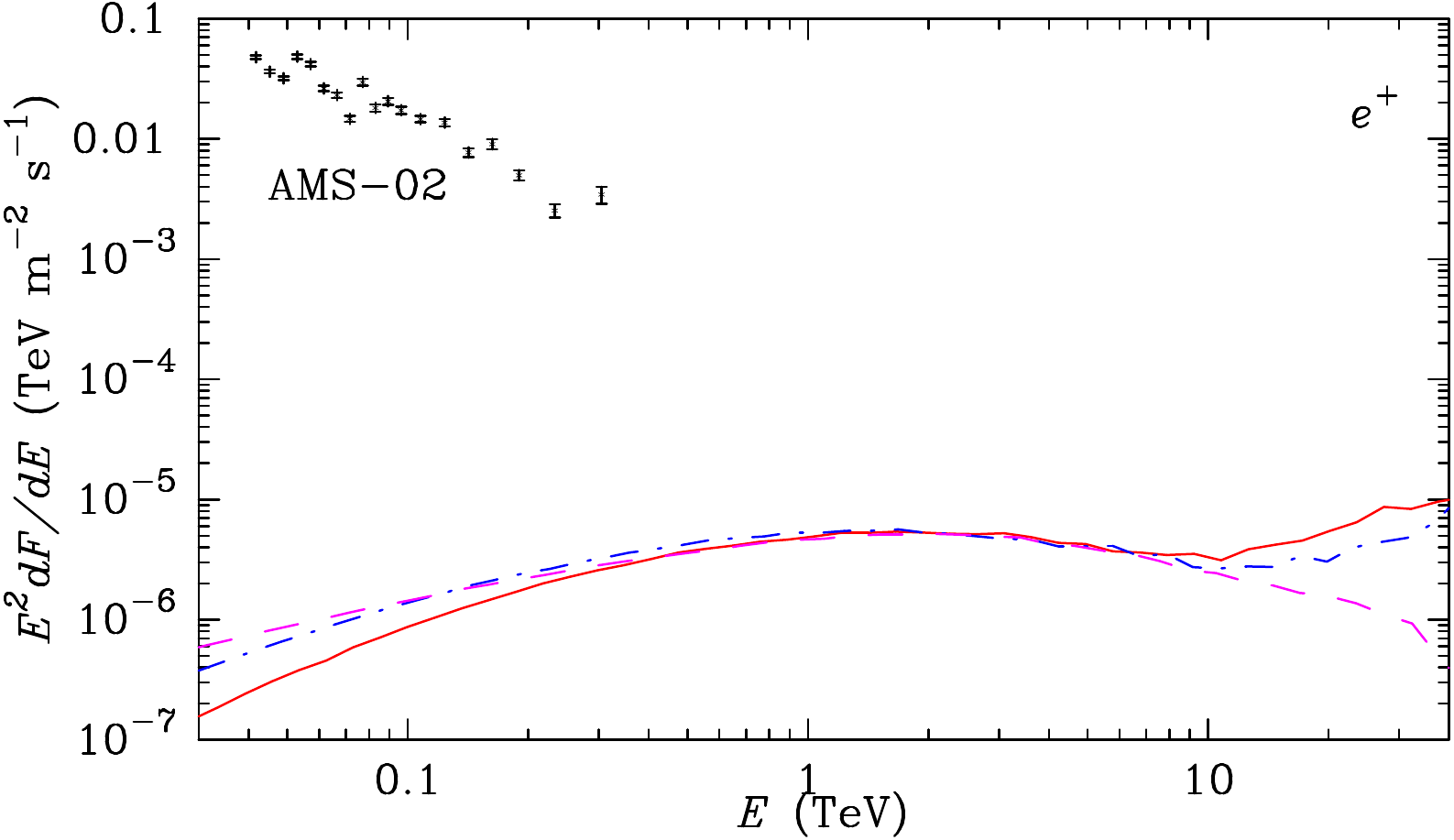} & \includegraphics[width=0.3\textwidth]{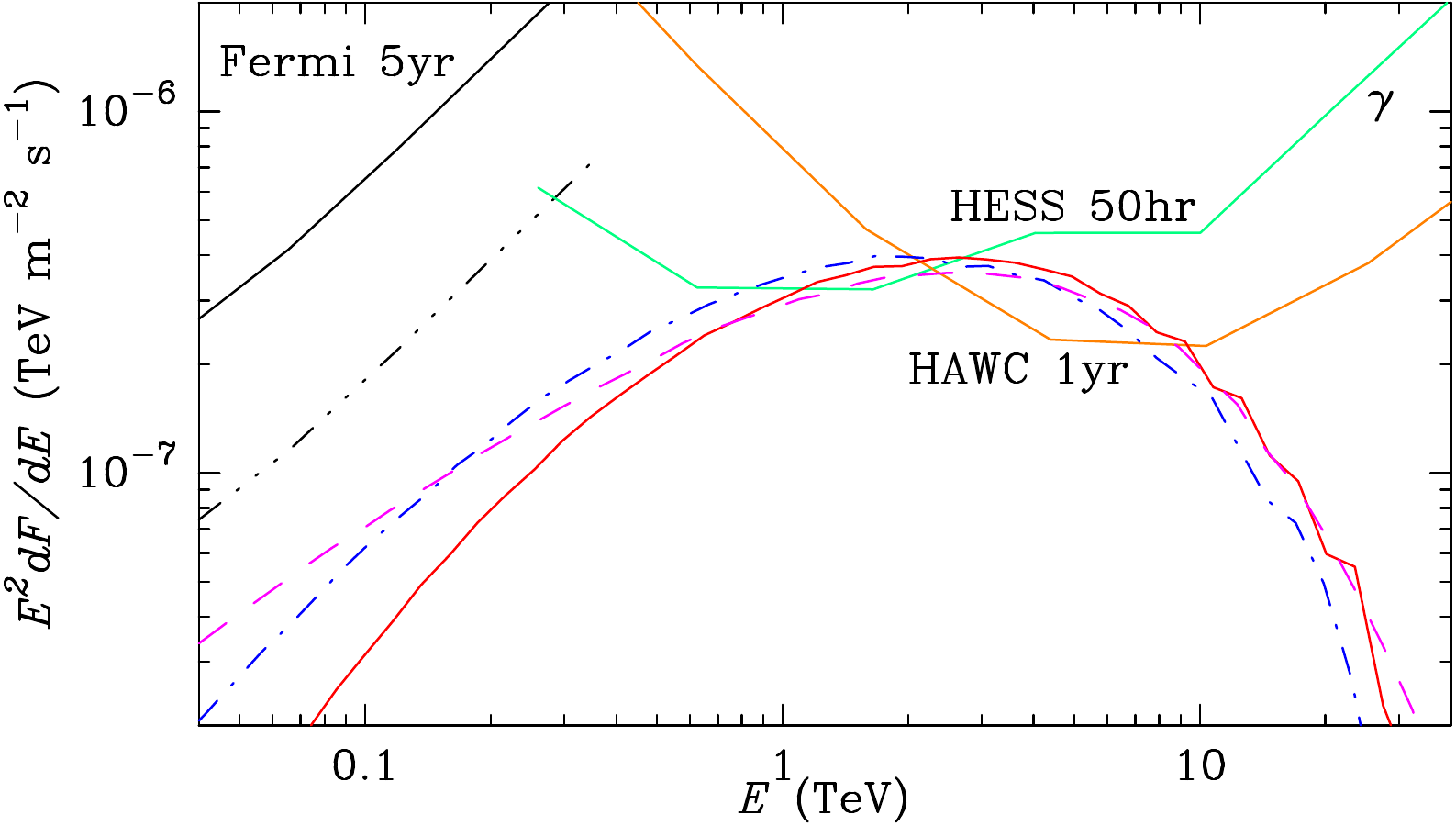}
\end{array}$
\caption[Figure of the limits on the DM CR model from several experiments.]{\label{limitsplots}: Figure of the limits on the three DM models of figure~\ref{DMCRspectra} from several experiments, for a $10^6M_{\astrosun}$ subhalo. For data sets which include a large number of data points, we checked the full spectrum for consistency with the model but only show the few highest-energy points. For isotropic limits, we compare the total flux from the DM subhalo with the total flux from the isotropic limit. 
For the gamma-ray figures, we show the gamma-ray sensitivities of three experiments to $5\degree$ extended sources (solid lines). The Fermi data are from Ref.~\cite{Funk:2012ca}, the HESS extended-source behavior is from Ref.~\cite{HESSweb}, and the  HESS point-source sensitivity and HAWC sensitivity are from Ref.~\cite{Abeysekara:2013tza}. The experiment sensitivities are given only out to $\sim3\degree$ half-width, so we have extrapolated the $5\degree$ values from the data. For the Fermi gamma-ray limit, the sensitivity to a $1\degree$ source is also shown as a triple-dot dashed line, for comparison.
}
\end{center}
\end{figure*}

{\parindent0pt\it Antiproton Constraints. }
In the energy ranges of our model, there are relevant CR antiproton upper bounds from two current experiments: ARGO-YBJ~\cite{Bartoli:2012qe} and PAMELA~\cite{Adriani:2010rc}. Figure~\ref{limitsplots} shows these limits and demonstrates that our models do not over-produce measured antiproton limits. The ARGO-YBJ flux limits were calculated by combining the ARGO $\bar{p}/p$ limits with the measured CR proton spectrum from CREAM~\cite{Yoon:2011aa}. The PAMELA limits were calculated by subtracting PAMELA's best-fit antiproton model from the PAMELA data and interpreting the resulting error bars as upper limits.
Both the antiproton limits are calculated assuming an isotropic antiproton flux and therefore provide only weak constraints for a small $0.03 {\rm\ sr}$ region of CRs. Additionally, it is likely that both the magnitude and spectrum of the CRs in the small DM CR region are significantly different from the average isotropic values over the whole sky. 

{\parindent0pt\it Electron and Positron Constraints. }
The measured CR emission is significantly more hadronic than leptonic or photonic~\cite{Abdo:2008kr}, so any model of the CR anisotropy must lack a strong leptonic signal. Above 1 TeV, the considered DM models produce approximately two electrons for each produced proton, and an anisotropic region which contains only 30\% hadrons would be inconsistent with the Milagro measurement. However, being much lighter than protons, electrons and positrons diffuse and undergo energy losses over shorter length scales. Therefore, it is likely that any electron and positron production by the DM annihilation would isotropize and shift to lower energy, either of which would make them indistinguishable from background.

It is also important to verify that the DM model of the anisotropy does not overproduce CR positron measurements. The AMS-02 collaboration has recently released measured $e^+/(e^++e^-)$ ratios up to 350 GeV~\cite{Aguilar:2013qda}, which we have turned into fluxes using the CR electron spectrum from the PAMELA collaboration~\cite{Adriani:2011xv}. A comparison of the AMS-02 values to those from our DM model, neglecting the positron diffusion and losses discussed above, is shown in figure~\ref{limitsplots}. The small $0.03 {\rm\ sr}$ region on the sky produces only a small fraction of the total CR positrons, and the constraints only weaken when energy-losses during propagation are included. This small flux also means that the DM positron flux cannot be the cause the full-sky positron excess seen in AMS-02, which must be coming from a different source.

{\parindent0pt\it Photon Constraints. }
The strongest constraint on the products of DM annihilation from a local subhalo is from the photons produced; if the DM source were bright enough, we would expect to have seen it in the gamma-ray sky, and no experiment has seen such a source coincident with Region A. However, the location of the gamma-ray counterpart to the CR excess can be far removed from Region A, due to possible bending in the magnetic stream. Also, because the gamma-rays do not interact with the magnetic stream, the relevant J-factor is one calculated with the true distance to the source. For our canonical $10^6M_{\astrosun}$ subhalo at $108 {\rm\ pc}$, the J-factor for gamma-rays is an order of magnitude lower than that for CRs, an effect that is stronger for smaller subhalos (see Table~\ref{DMJtable}).
Finally, the sensitivities of the Fermi-LAT, HESS, and HAWC for a $5\degree$ half-angle source are over an order-of-magnitude worse than the corresponding point-source values~\cite{Funk:2012ca,HESSweb,Abeysekara:2013tza}.

From figure~\ref{limitsplots}, we can see that the Fermi-LAT will not have detected the DM source of the CR excess in its 5 years of operations, even if the source were extended by only $1\degree$ half-width. The DM gamma rays should be detectable by the imaging air Cherenkov telescopes (IACTs), however, despite the large source extent. The caveat is that IACTs have a very small field-of-view and would likely only observe the source as a follow-up to a marginal detection by the Fermi-LAT.

The HAWC observatory will have a similar sensitivity to the IACTs and a wide field-of-view. After less than 1 year of data collection, HAWC will be able to verify the gamma-ray emission of a bright nearby DM subhalo if it is in HAWC's large field-of-view of $\sim 8{\rm\ sr}$. The Milagro observatory was a wide field-of-view TeV gamma-ray observatory similar to HAWC but with $\sim15$ times less sensitivity. Especially if the DM source has a latitude much lower than Region A, Milagro would not have observed the DM source in gamma-rays. For all Earth-based detectors, it is possible that the DM subhalo is located at an angle outside of the detector's view. Detectors also may not view the DM subhalo in gamma-rays if it is situated between the Sun and the Galactic center (GC)~\cite{Vincent:2010kv} or if the magnetic stream had a focusing effect and the source is larger than $5\degree$.

{\parindent0pt\it The Dark Matter Cross-Section. }
For standard thermal relic DM, a cross-section of $\langle\sigma_{\rm{A}}v\rangle\approx3\times10^{-26}\rm\ cm^3\,s^{-1}$ in the early universe is needed in order to produce the DM density observed today. However, the DM cross-section today may be much higher. If the DM  couples to a boson which is much lighter than the DM mass, it can create a resonance which increases the DM cross-section over thermal at low velocities, called ``Sommerfeld enhancement''. For multi-TeV DM masses, standard-model gauge bosons are light enough to cause this Sommerfeld enhancement. This ``natural'' Sommerfeld enhancement occurs for the $Z^0Z^0$ annihilation channel or in the case of the annihilation taking place though a $Z^0$ boson (i.e. $\chi\chi\rightarrow Z^0\rightarrow W^+W^-$). The natural Sommerfeld enhancement for the three considered DM models are shown in Table~\ref{Sommtable}, calculated using the formalism of Ref.~\cite{Feng:2010zp}, with $M_{Z^0}=91.2 {\rm\ GeV}$ and a weak coupling of $\sim1/35$.

Aside from Sommerfeld enhancement, details of the DM spatial distribution and the DM particle model can increase the flux as well. DM substructure can both increase the Sommerfeld enhancement and J-factor of the subhalo~\cite{Slatyer:2011kg}. If the subhalo were very close to the magnetic stream, a greater fraction of the emitted flux could be propagated to Earth, increasing the CR flux.
Also, in the case of non-thermal DM models or DM models with light dark-sector mediators, the DM cross-section can be much larger than thermal.

A $10^6M_{\astrosun}$ DM subhalo requires a DM cross-section $\langle\sigma_{\rm{A}}v\rangle\approx6\times10^{-23}\rm\ cm^3\,s^{-1}$, which is currently allowed by the VERITAS Segue 1 limits on the DM cross-section~\cite{Aliu:2012ga}. Though the HESS GC limits are much stronger, Ref.~\cite{Slatyer:2011kg} demonstrated that limits from the GC cannot be directly applied to local DM subhalos which include Sommerfeld enhancement. Therefore, our DM models, which have an un-enhanced cross-section less than $2\times10^{-24}\rm\ cm^3\,s^{-1}$, are consistent with the HESS GC limits on DM~\cite{Abramowski:2011hc,Abazajian:2011ak}.
Furthermore, a 50 TeV $Z^0Z^0$-channel DM annihilating in a nearby $10^6M_{\astrosun}$ subhalo with a natural Sommerfeld-enhanced thermal cross-section produces the measured CR anisotropy, without the need for any other astrophysical boosts or non-standard DM models.
\begin{table}
\begin{center}
\begin{tabular}[t]{|c|c|c|c|}
  \hline
   & $60 {\rm\ TeV}\ W^+W^-$ & $50 {\rm\ TeV}\ Z^0Z^0$ & $100 {\rm\ TeV}\ b\bar{b}$ \\
  \hline
  Sommerfeld & 130 & 1301 & 226 \\
  \hline
\end{tabular}
\caption[Table of the natural Sommerfeld-enhanced cross-sections for the DM candidates.]{\label{Sommtable}: Table of the natural Sommerfeld-enhanced cross-sections for the DM candidates. The Sommerfeld enhancement is calculated assuming a $v_{\rm rel}$ of $10 {\rm\ km/s}$ in the subhalo and an annihilation channel through $Z^0$ bosons. In the $Z^0Z^0$ case, the $50 {\rm\ TeV}$ mass lies near a resonance at $47 {\rm\ TeV}$, so the Sommerfeld enhancement is greater, though not by the orders of magnitude a fine-tuned mass would give.}
\end{center}
\end{table}

{\parindent0pt\it Discussion. }
Annihilating TeV DM in a nearby subhalo is a strong candidate for the source of the observed TeV CR excess. A DM subhalo connected to the Sun by a magnetic stream naturally explains the spectrum of the CR anisotropy and the lack of a corresponding gamma-ray signal. Additionally, we have shown that signal is not strongly dependent on the DM particle mass, the DM spectrum, or the mass of the DM subhalo. In our analysis, we have chosen conservative values for the magnetic stream and the size of the DM subhalo, and we have neglected all focusing effects of the magnetic field as well as any DM substructure. With our conservative choice of DM model, we have shown that it is possible to naturally produce the CR anisotropy from DM annihilation without violating any current observations. Further analysis of the details of the local magnetic field and the expected HAWC CR anisotropy measurement should improve our understanding of the DM source of the CR excess as well.

If DM annihilations are the source of the CR anisotropy, however, then many CR discoveries are just around the corner. AMS-02 could detect a positron excess with a hard spectrum in its pointed data near Region A. An IACT follow-up to a Fermi-LAT hard-spectrum source would readily detect a DM source emitting 1 Crab of gamma-rays. Also, HAWC should observe the DM subhalo as an extremely bright, extended gamma-ray source. The next few years will be exciting for TeV cosmic rays.
%
\begin{acknowledgments}
{\parindent0pt\it Acknowledgments. }
We thank Kevork Abazajian, Brenda Dingus, Peter Karn, Gerd Kunde, John Pretz, and Patrick Younk for useful discussions. JPH is supported by the Los Alamos National Laboratory and the Department of Energy.
\end{acknowledgments}
%
\bibliography{bibliography}
\end{document}